\documentclass[twocolumn,showpacs,prb,amsmath,amssymb,aps,superscriptaddress]{revtex4-2}
\usepackage{dcolumn}
\usepackage{autobreak}
\usepackage{braket}
\usepackage{graphicx}
\usepackage{booktabs}
\usepackage{lipsum}
\usepackage{bm}
\usepackage[colorlinks, linkcolor=blue, anchorcolor=blue, citecolor=blue]{hyperref}  

\makeatletter

\newcommand{\Rmnum}[1]{\expandafter\@slowromancap\romannumeral #1@}
\newcommand{\SRO}{Sr$_2$RuO$_4$}

\newcommand{\chiralp}{$p_x+ip_y$ }
\newcommand{\pd}{\partial}
\makeatother

\usepackage{mynote}

\begin{document}
\title{Examining the possibility of chiral superconductivity in \SRO~and other compounds via applied supercurrent}
\author{Hao-Tian Liu}
\affiliation{Shenzhen Institute for Quantum Science and Engineering, Southern University of Science and Technology, Shenzhen 518055, Guangdong, China}
\affiliation{International Quantum Academy, Shenzhen 518048, China}
\affiliation{Guangdong Provincial Key Laboratory of Quantum Science and Engineering, Southern University of Science and Technology, Shenzhen 518055, China}
\author{Weipeng Chen}
\affiliation{Shenzhen Institute for Quantum Science and Engineering, Southern University of Science and Technology, Shenzhen 518055, Guangdong, China}
\affiliation{International Quantum Academy, Shenzhen 518048, China}
\affiliation{Guangdong Provincial Key Laboratory of Quantum Science and Engineering, Southern University of Science and Technology, Shenzhen 518055, China}
\author{Jia-Xin Yin}
\affiliation{Department of Physics, Southern University of Science and Technology, Shenzhen, Guangdong, China}
\author{Cai Liu}
\affiliation{College of Integrated Circuits and Optoelectronic Chips, Shenzhen Technology University, Shenzhen 518118, Guangdong, China}
\author{Wen Huang}
\email{huangw3@sustech.edu.cn}
\affiliation{Shenzhen Institute for Quantum Science and Engineering, Southern University of Science and Technology, Shenzhen 518055, Guangdong, China}
\affiliation{International Quantum Academy, Shenzhen 518048, China}
\affiliation{Guangdong Provincial Key Laboratory of Quantum Science and Engineering, Southern University of Science and Technology, Shenzhen 518055, China}

\begin{abstract}
One approach to probe the still controversial superconductivity in \SRO~is to apply external perturbations that break the underlying tetragonal crystalline symmetry. Chiral $p_x+ip_y$ and $d_{xz}+id_{yz}$ states respond to such perturbations in ways that may help to distinguish them from other superconducting pairings. However, past experimental efforts along this line, using uniaxial strains and magnetic fields parallel to the RuO$_2$ plane, have not been able to reach an unambiguous conclusion. In this study, we propose to further examine the possibility of chiral superconducting order in \SRO~using an alternative tetragonal-symmetry-breaking perturbation --- in-plane supercurrent. We study the superconducting phase diagram as a function of both temperature and the applied supercurrent. Supercurrent generically splits the transition of the two chiral order parameter components, and we show that the splitting can give rise to visible specific heat anomalies. Furthermore, supercurrent parallel and anti-parallel to the unidirectional propagation of the chiral edge modes impact the edge states in different manner. This difference manifests in the tunneling spectrum, thereby providing an additional means to probe the chirality even when the related spontaneous edge current is vanishingly small. Finally, we discuss the distinction of supercurrent responses in non-chiral time-reversal-symmetry-breaking superconducting states. Our proposal can be applied to other candidate chiral superconductors. 

\end{abstract}

\date{\today}

\maketitle
\section{Introduction}

The Cooper pairing in chiral superconductors spontaneously breaks time-reversal symmetry, which can be detected by zero-field muon spin relaxation and optical polar Kerr effect measurements. Intrinsic chiral superconductors are hard to come by. Existing signatures consistent with chiral states have only been reported in a limited few materials, including \SRO~\cite{Luke1998, Xia2006}, UPt$_3$~\cite{Luke1993, Schemm2014} and URu$_2$Si$_2$~\cite{Mac1988, Schemm2015}, although none of the above is unambiguously confirmed chiral superconductor. 

\SRO~\cite{Maeno1994} is among the most thoroughly studied candidate chiral superconductor. In the past three decades, much progress has been made towards unraveling the myth of its superconductivity. However, the exact nature of the Cooper pairing in this material remains hugely controversial~\cite{Maeno2001,Maeno2003,Maeno2003, Kallin2009, Kallin2012, Maeno2012, Liu2015, Kallin2016, Mac2017, Huang2021CPB, Leggett2021}. In particular, while multiple early observations pointed to chiral $p$-wave pairing~\cite{Luke1998, Ishida1998, Duffy2000, Liu2004,Xia2006}, a number of recent key observations seem to defy a straightforward chiral $p$-wave interpretation~\cite{Pus2019, Chronister2021}. It is worth stressing that, while many contending candidate pairing symmetries have recently emerged in conjunction with numerous experimental advances~\cite{Pus2019, Chronister2021,exp1,exp2,exp3,ultrasound1,ultrasound2}, no order parameter seems able to coherently interpret all of the key observations. Nonetheless, chiral superconducting states such as $p+ip$ or $d+id$ cannot be ruled out at this stage, as they may be crucial to explain the polar Kerr effect~\cite{Huang2021CPB,ZhangJL:20,LiuHT:2023}. It is important to stress that, multi-component states such as $s+id_{x^2-y^2}$ and $d_{x^2-y^2}+ig_{xy(x^2-y^2)}$ preserve certain vertical mirror symmetry~\cite{Kivelson2020} and are thus incompatible with the Kerr rotation. This holds even in the presence of random disorder~\cite{LiuHT:2023}.

\begin{figure}
\includegraphics[width=8.5cm]{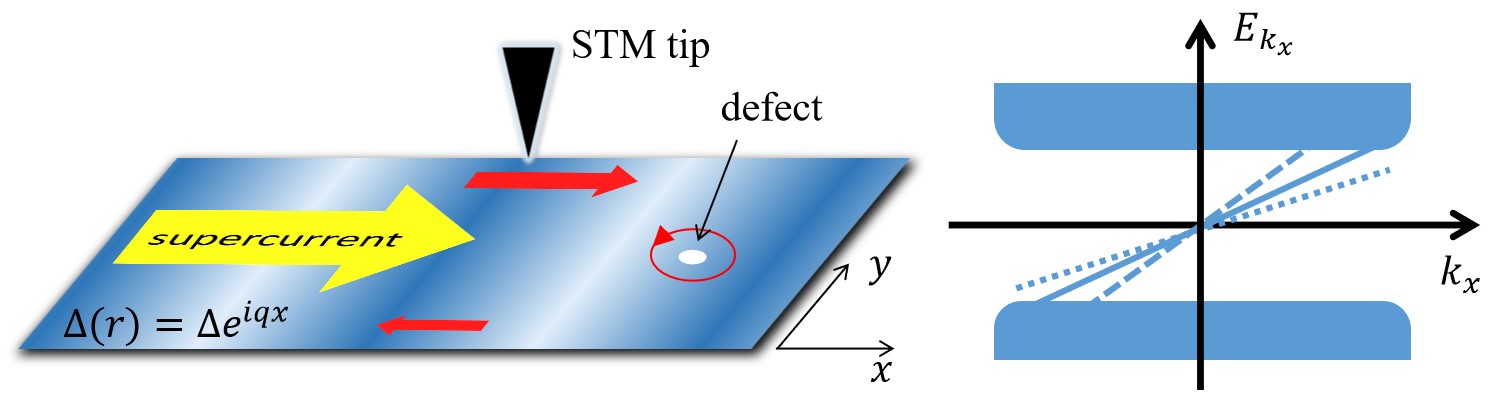}
\caption{Left: Sketch of a two-dimensional chiral superconductor in the presence of an injected supercurrent (yellow arrow). The red arrows indicate the flow of spontaneous edge current or the group velocity of chiral edge modes. The different size of the two red arrows reflects the unequal supercurrent-induced changes to the spontaneous edge current and chiral edge dispersion. Right: Sketch of the low-energy quasiparticle spectrum at one of the edges parallel to the $x$-direction. The in-gap chiral edge dispersion for cases without supercurrent, with left and right supercurrent are displayed as solid, dashed and dotted lines, respectively. Note the continuum spectrum in the presence of supercurrent is not sketched. The supercurrent-induced low-energy spectral variation at any boundary can be probed by tunneling spectroscopic techniques, such as STM and point contact. Similar asymmetric response to opposite supercurrents is also expected in the vicinity of impurities, defects, etc (shown on left panel).}
\label{fig0}
\end{figure}

One approach to identify chiral pairing (e.g. $p_x+ip_y$, $d_{xz}+id_{yz}$) is to apply a perturbation that affects the $p_x$ ($d_{xz}$) and $p_y$ ($d_{yz}$) components of the superconducting order parameter in different manner~\cite{Sigrist1991}. For example, in-plane uniaxial strain breaks the tetragonal symmetry of \SRO~and may therefore lift the degeneracy between the two components~\cite{Hicks2014}. However, the conclusion drawn from strain measurements is not conclusive. On the one hand, muon spin relaxation under strain indicates a splitting between an upper superconducting transition and a lower one breaking time-reversal symmetry~\cite{Grinenko2021}. On the other hand, dedicated thermodynamic study has failed to observe any signature ascribable to two successive transitions~\cite{Li2021,Mueller:23}, while a number of other measurements did not reveal the expected linear cusp associated with the purported upper transition~\cite{Hicks2014, Steppke2017,Watson2018}. Apart from in-plane strain, similar symmetry breaking is also achieved with the application of in-plane magnetic field~\cite{Agterberg:98,Kaur:05}. Interestingly, once again no clear secondary transition has been observed in the most up-to-date specific heat experiment under in-plane fields~\cite{Yonezawa2014}. 

Another analogous idea is to inject a supercurrent which can also lift the degeneracy of the two chiral components. Several previous literature have studied the tunneling spectrum of a single-component superconductor subject to a supercurrent~\cite{Anthore:03,Zhang2004} and the supercurrent-induced phase transitions between different pairing states in multiband $s$-wave superconductors~\cite{Yer2022,Yer2023}. In this work, we study the effect of in-plane d.c.~supercurrent on chiral superconducting states. While our calculations will be based on a $p_x+ip_y$ state, the conclusions can be generalized to $d_{xz}+id_{yz}$ and other chiral states. Compared to the unaxial strain, supercurrent bias may be a more advantageous perturbing probe, as the former may suffer from strain inhomogeneity that could hinder the identification of signatures associated with chiral pairing.

Generally speaking, the leading order impact of a supercurrent $J$ is captured by the following terms in the free energy (assuming a charge neutral system),
\begin{equation}
(\alpha + a_x J^2)|\Delta_x|^2 + (\alpha + a_y J^2)|\Delta_y|^2 \,.
\label{eq:Jcoupling}
\end{equation}
Here, $\Delta_{x/y}$ denote the two chiral order parameter components, $\alpha$ and $a_{x/y}$ are coefficients that depend on microscopic details. Note that the absence of terms linear in $J$ is because supercurrent flowing parallel and antiparallel to a specific direction should have the same effect on any individual order parameter component in the bulk. Hence, unlike the theoretically expected linear variation of the transition temperature with uniaxial strain, the critical temperature $T_c$ should follow a quadratic dependence on small supercurrent, as will be verified in our calculations (Fig.~\ref{fig3}). 

We shall model the current-carrying state by considering Cooper pairings that exhibit finite center-of-mass momentum $2\bq$, i.e.~a Fulde-Ferrell state~\cite{FF1964} (with supercurrent $J \propto \bq$). By means of self-consistent mean-field Bogoliubov de-Gennes calculations, we then determine the phase diagram as a function of both $\bq$ and temperature. We further show that the splitting of the two chiral components will also manifest as two anomalies in specific heat. 

Besides the bulk probe mentioned above, chiral edge modes and spontaneous edge current are another set of phenomena often associated with chiral superconductivity~\cite{Stone2004}. However, while the chiral edge modes are topologically protected, the edge current is not~\cite{Huang2015, Tada2015}. The latter can be sensitive to microscopic details and disorder~\cite{Imai201213,Huang2014,Lederer2014,Scaffidi2015, Tada2015b, Huang2015}, prompting an argument that the experimental null result on the edge current in \SRO~\cite{Kirtley2007, Hicks2010, Curran2014} can still be compatible with a chiral superconducting state. Hence edge current may not serve as an effective diagnosis of chiral superconductivity. We propose in this study an alternative diagnosis by looking at the supercurrent-induced changes to the chiral edge modes: at any edge of a chiral superconductor, left and right supercurrents perturb the chiral edge dispersion differently, giving rise to different corrections to the edge tunneling spectrum (see right panel of Fig.~\ref{fig0}). By simple extension, similar asymmetric supercurrent-induced correction is also expected in the vicinity of other forms of translation symmetry breaking perturbations, such as impurities and defects, around which circulating spontaneous supercurrent emerge (see Fig.~\ref{fig0}, left panel). 

The rest of the paper is organized as follows. In Sec.~\ref{sec:formalism} we introduce the self-consistent mean-field formalism for studying the effects of supercurrent. The numerical results are presented in Sec.~\ref{sec:results}. Section \ref{subsec2} presents the $q$-$T$ phase diagram, and Sec.~\ref{subsec3} demonstrates the supercurrent-induced splitting of the specific heat anomaly. While in Sec. \ref{subsec4}, we focus on the response of chiral edge modes to supercurrent and study the distinct changes to the edge tunneling spectrum in the presence of opposite supercurrents. Section \ref{sec3} concludes the paper by discussing the identification of other time-reversal-symmetry-breaking (TRSB) superconducting states by means of applying a supercurrent.

\section{Formalism}
\label{sec:formalism}
In the absence of supercurrent, Cooper pairing takes place between electrons with opposite momenta $\bk$ and $-\bk$. A supercurrent state is described by Cooper pairings with a net center-of-mass momentum $2\bq$, i.e. between electrons with momenta $\bk+\bq$ and $-\bk + \bq$. Equivalently, the superconducting order parameter acquires real-space phase modulation $\Delta \ra \Delta e^{2i\bq\cdot \bR}$~\cite{Tinkham2004}, where $\bR$ is the center-of-mass position of the Cooper pair. Consider a simplified single-band spinless \chiralp model on a square lattice, an effective Hamiltonian then follows as:
\begin{align}
    \hH = \sum_{\bk}\xi_{\bk}\hcd_{\bk}\hc_{\bk}+\sum_{\bk,\bk'}\widetilde{U}_{\bk,\bk'}\hcd_{\bk+\bq}\hcd_{-\bk+\bq}\hc_{-\bk'+\bq}\hc_{\bk'+\bq} \,,
\end{align}
where $\xi_{\bk}=2t(\cos k_x+\cos k_y)-\mu$ is the electron dispersion in the normal state, $t$ is the hopping amplitude, $\mu$ is the chemical potential, and $\widetilde{U}_{\bk,\bk'}$ is the effective interaction in the p-wave Cooper channel. The effective interaction can be decomposed as $\widetilde{U}_{\bk,\bk'}=\widetilde{U}_{x,\bk,\bk'}+\widetilde{U}_{y,\bk,\bk'}$, where the two terms on the right hand side are to generate $p_x$ and $p_y$ pairings respectively and each can be written in a separable form $\widetilde{U}_{\alpha,\bk,\bk'}=-\widetilde{U}_0f_{\alpha,\bk}f^*_{\alpha,\bk'}$ ($\alpha = x, y$). Here $\widetilde{U}_0$ denotes the strength of the effective attraction and $f_{\alpha,\bk}$ represents the form factor of the $p_\alpha$-wave pairing. 

Then the chiral p-wave gap function $\Delta_{\bk} = \Delta_x \sin k_x + \Delta_y \sin k_y$ is determined self-consistently with initial values $(\Delta_x, \Delta_y) = (1, i)\Delta_0$. The self-consistent expression is given by:
\begin{align}\label{selfc}
    \Delta_{\bk}(\bq, T) = \frac{1}{N}\sum_{\bk'} \widetilde{U}_{\bk, \bk'} \langle \hat{c}_{-\bk' + \bq} \hat{c}_{\bk' + \bq} \rangle
\end{align}
where $N$ is the number of unit cells, and $\langle \cdots \rangle$ denotes the expectation value in the ground state. By utilizing \eqref{selfc}, we are able to examine how the amplitude of the order parameter, denoted as $\Delta_x$ and $\Delta_y$, varies with temperature and supercurrent. The expression for these variations is determined by:
\begin{align}
    \Delta_{\alpha}(\bq, T) = \frac{1}{N}\sum_{\bk} \widetilde{U}_{\alpha} f_{\alpha} \langle \hat{c}_{-\bk + \bq} \hat{c}_{\bk + \bq} \rangle
\end{align}
Here, $f_{\alpha} = \sin k_{\alpha}$ for $p_{\alpha}$-wave pairing, and we set $\widetilde{U}_x = \widetilde{U}_y = \widetilde{U}_0$. Different phases can be determined based on these variations.

Furthermore, the mean-field Hamiltonian is given by:
\begin{align}\label{mf_hamiltonian}
    \hat{H} = &\sum_{\bk} \xi_{\bk + \bq} \hat{c}^{\dagger}_{\bk + \bq} \hat{c}_{\bk + \bq} \nonumber \\
    &+ \sum_{\bk} (\Delta_{\bk} \hat{c}^{\dagger}_{\bk + \bq} \hat{c}^{\dagger}_{-\bk + \bq} + \text{h.c.}) + \frac{N|\Delta_{\bk}|^2}{\widetilde{U}_0}
\end{align}
and it can be also expressed by a matrix in the Nambu basis $\Psi_{\bk}=(\hc_{\bk+\bq}, \hcd_{-\bk+\bq})^t$ as follows:
\begin{align}\label{BdG_matrix}
\mathcal{H}_{\bk}(\bq)=\begin{pmatrix}
\xi_{\bk+\bq} & \Delta_{\bk}\\
\Delta_{\bk}^* & -\xi_{-\bk+\bq}
\end{pmatrix}
\end{align}
Hence, the quasiparticle energy can be derived from the Hamiltonian as:
\begin{align}\label{energy}
    E_{\bk,\pm}(\bq) &= \frac{1}{2}(\xi_{\bk + \bq} - \xi_{-\bk + \bq}) \nonumber \\
    &\pm \sqrt{\bigg[\frac{1}{2}(\xi_{\bk + \bq} + \xi_{-\bk + \bq})\bigg]^2 + |\Delta_{\bk}|^2}
\end{align}
At small $\bq$, the above expression can be approximated as $E_{\bk,\pm}(\bq) \simeq \bk\cdot \bq/m \pm \sqrt{\xi_{\bk}^2 + |\Delta_{\bk}|^2} \simeq  \boldsymbol{v}_F\cdot \bq \pm \sqrt{\xi_{\bk}^2 + |\Delta_{\bk}|^2}$. This dispersion includes a Doppler shift in the continuum limit~\cite{Volovik1993}, and it is no longer symmetric along $E=0$ due to the presence of a supercurrent. 

Through out the study, we do not take into account the screening effects in a charged superfluid. However, all of the major conclusions are expected to uphold qualitatively.

\section{Numerical Results}
\label{sec:results}
\subsection{Phase diagram}
\label{subsec2}
To simplify notation, we take $\hbar = c = k_B = e=1$. Through all calculations, we assume $t=1$ and take the supercurrent to flow along $x$ direction, $\bq = (q_x,0)$. 

\begin{figure}
  \centering
  \includegraphics[width=1\linewidth]{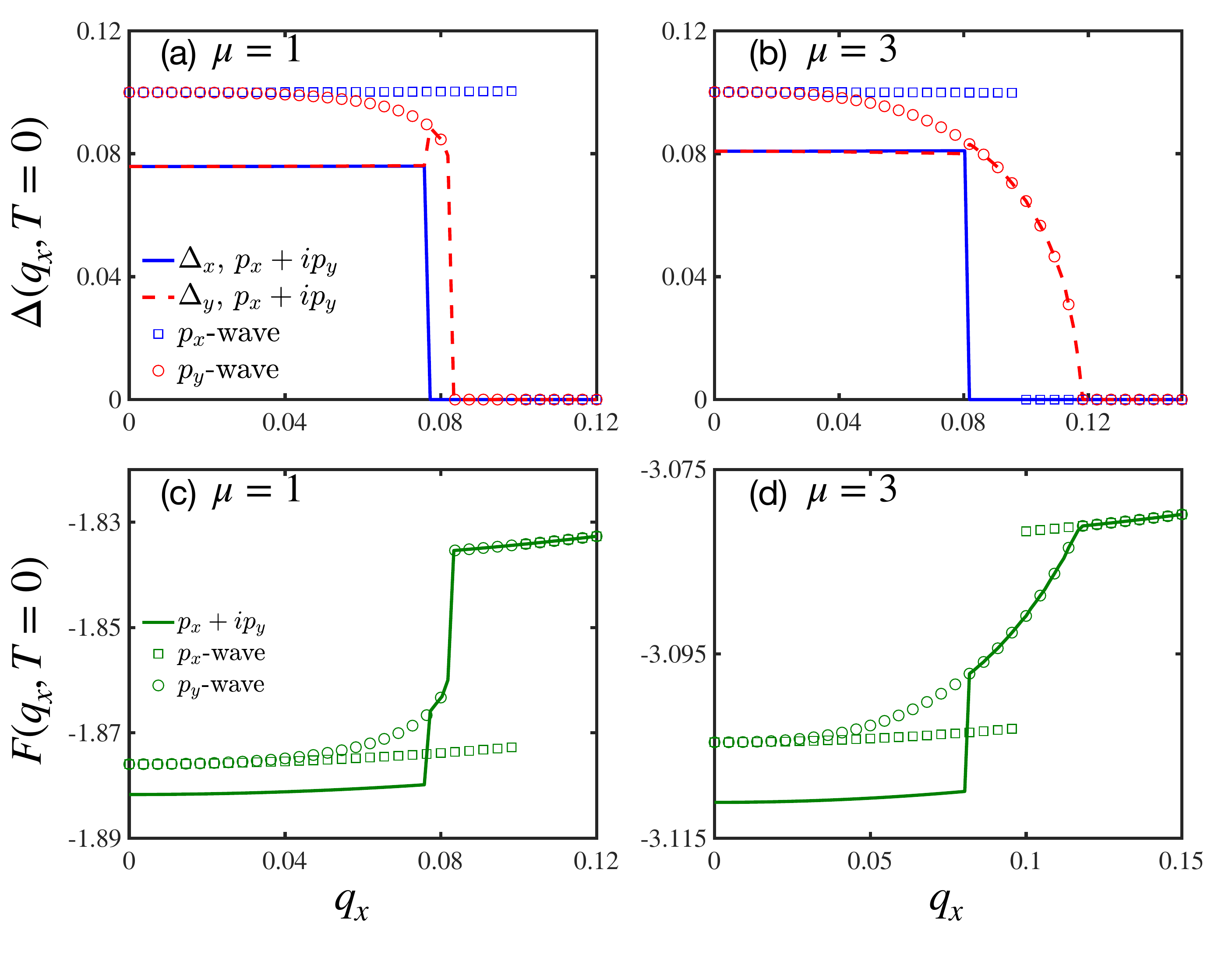}
  \caption{The variation of self-consistent order parameters (upper panel) and free energy (lower panel) as a function of $q_x$ at zero temperature. Note that the label in each case denotes the corresponding initial order parameter configuration, which does not necessarily coincide with the final self-consistent configuration. The reference energy scale is chosen to be $t=1$. The left and right panels show results for different chemical potentials, $\mu=1$ and $\mu=3$ respectively. In each case the strength of the pairing interaction $\tilde{U}_0$ is chosen such that the self-consistent pairing amplitude $\Delta_0 = 0.1$ in the pure $p_x$- or $p_y$-wave states. These calculations are performed in the k-space with a size of $N=1000\times 1000$. }
  \label{fig1}
\end{figure}

\begin{figure}
  \centering
  \includegraphics[width=1\linewidth]{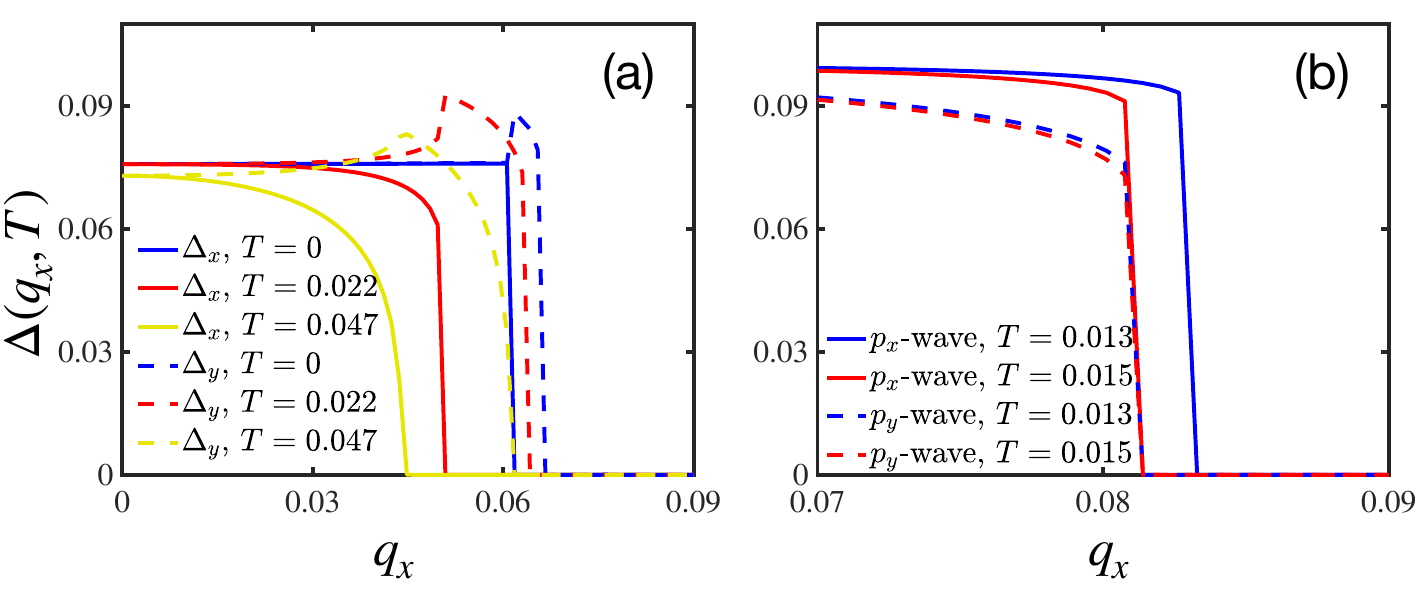}
  \caption{The evolution of self-consistent order parameters as a function of $q_x$ at different temperatures for calculations with different initial superconducting order parameter configurations: (a) \chiralp and (b) $p_{x/y}$-wave. The chemical potential is set at $\mu=1$, while other parameters are the same as in Fig. \ref{fig1} (a).}
  \label{fig2}
\end{figure}

Figure~\ref{fig1} (a) and (b) shows a representative set of zero-temperature results for $p_x$-wave, $p_y$-wave and $p_x+ip_y$ pairings. One noteworthy feature is the distinct response of the order parameter in the single-component $p_x$ and $p_y$-wave states to supercurrent. In particular, while the $p_x$-wave order parameter $\Delta_x$ barely changes with increasing $q_x$ prior to a first order transition beyond which it vanishes, $\Delta_y$ varies more drastically as a function of $q_x$. This distinction can be attributed to the rather different supercurrent-induced correction to the quasiparticle spectrum, which happens in conjunction with the different nodal position of the individual gap functions~\cite{Zhang2004}. The $p_x$ pairing has its nodal points at $\bk=(0,\pm k_F)$, hence the Doppler shift at the nodal momenta is $\delta E_{\bk} \approx \bk\cdot \bq/m = 0 $. For $p_y$ pairing, the nodal points are located at $\bk=(\pm k_F, 0)$, hence $\delta E_{\bk} \approx k_Fq_x/m$. At the special chemical potential $\mu=0$ in our particle-hole symmetric square lattice model, the four nodal momenta of $p_x$ and $p_y$ pairings coincide. We have checked that the two order parameters vary in very much similar fashion in that special case. The above behavior of $\Delta_x$ and $\Delta_y$ does not necessarily carry over to the two-component $p_x+ip_y$ calculations at finite temperature, due to the change of gap structure. For example, at low but finite temperatures and small $q_x$, $\Delta_x$ in the $p_x+ip_y$ state could change more rapidly than $\Delta_y$ does as a function of $q_x$ (Fig. \ref{fig2} (a)), unlike in the single-component calculations (Fig.~\ref{fig2} (b)).

\begin{figure}
  \centering
  \includegraphics[width=1\linewidth]{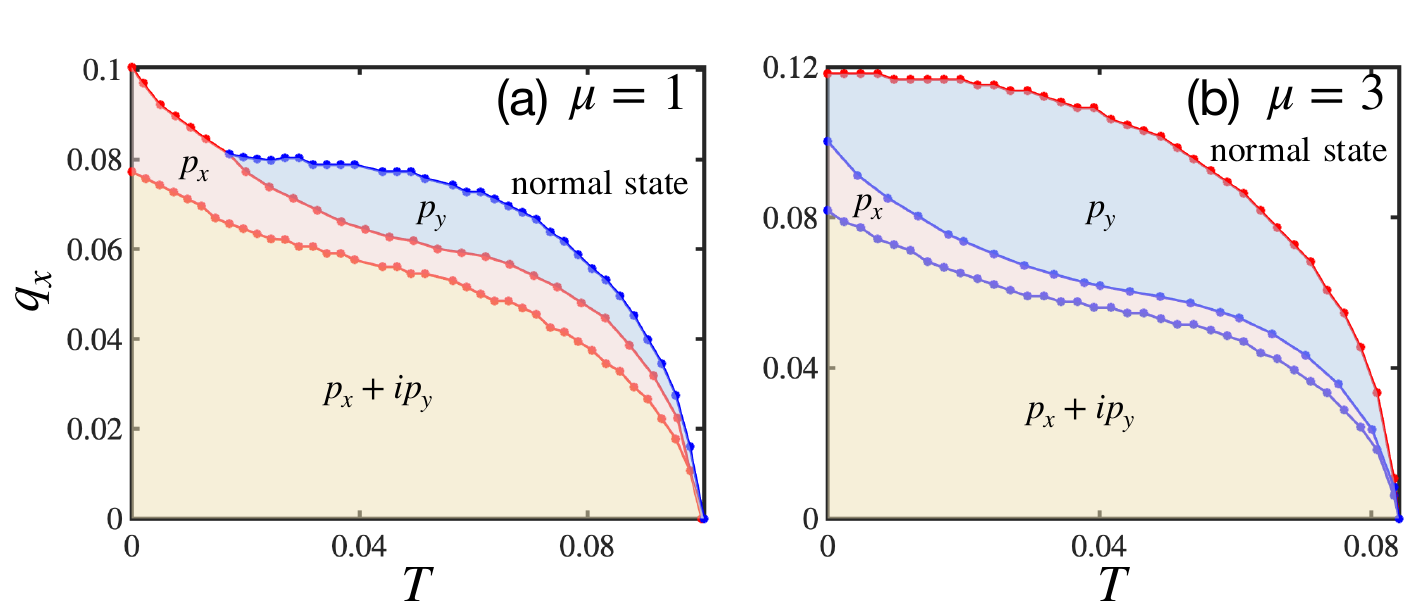}
  \caption{The $q_x$-$T$ phase diagram determined by free energy calculation at chemical potentials (a) $\mu=1$ and (b) $\mu=3$. Parameters are the same as in Fig. \ref{fig1}. }
  \label{fig3}
\end{figure}

In the presence of a supercurrent, it is not known apriori whether chiral p-wave will be energetically more favorable than the nonchiral single-component $p_x$-wave and $p_y$-wave pairings. Furthermore, it is unclear whether self-consistent iterations with an initial chiral p-wave pairing condition can return the correct lowest energy state at large supercurrent. This can be seen in Fig.~\ref{fig1} (a) and (b). On the one hand, the relative robustness of single-component $p_x$ ($\Delta_x$) and $p_y$ ($\Delta_y$) pairings against supercurrent varies with model detail, i.e. chemical potential: at $\mu=1$, $\Delta_x$ persists to stronger supercurrent, and it is the other way around at $\mu=3$. On the other hand, in chiral p-wave calculations, $\Delta_y$ in general seems to survive up to stronger supercurrent than $\Delta_x$ does. Hence self-consistent calculations with initial chiral p-wave condition may in some circumstances obtain states that are not lowest energy, as exemplified in Fig.~\ref{fig1} (c). We thus consider all three different initial conditions, and compare their final self-consistent free energy $F(q_x,T)=\braket{\hH}$ to determine the phase diagram. The $q_x$-$T$ phase diagrams of the above calculations are shown in Fig. \ref{fig3}. At finite $q_x$, the two order parameter components generically onsets at different temperatures. In most regime, $\Delta_y$ is more robust against supercurrent, followed by a pure $\Delta_x$ state at lower $q_x$, before eventually giving way to the two-component chiral $p$-wave state at even lower $q_x$. However, due to the above-mentioned subtle dependence on microscopic parameters, in the case of $\mu=1$ and at low temperatures and large $q_x$, $\Delta_x$ persists to higher $q_x$ and no pure $\Delta_y$ exists (Fig. \ref{fig3} (a)). 

\subsection{Specific heat}
\label{subsec3}
\begin{figure}
  \centering
  \includegraphics[width=1\linewidth]{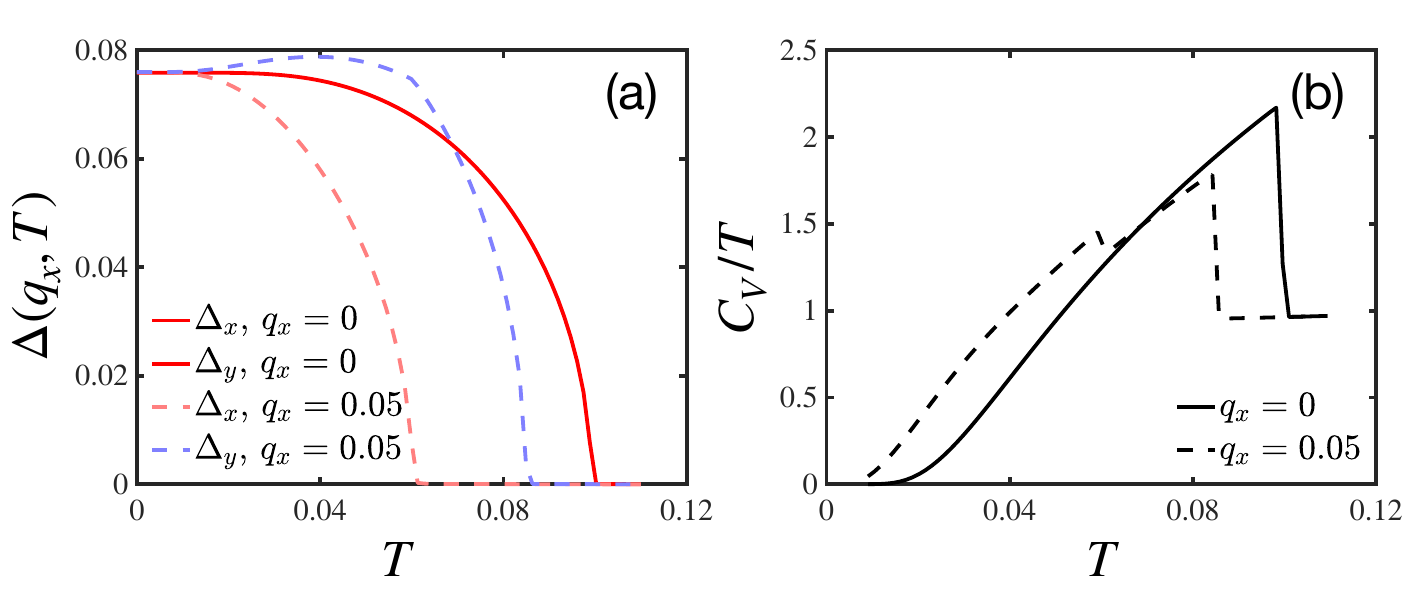}
  \caption{The evolution of self-consistent order parameters (a) and the specific heat (b) as functions of temperature in the chiral p-wave model with $\mu=1$, considering different supercurrent momenta. Parameters are the same as in Fig. \ref{fig1}. }
  \label{fig4}
\end{figure}

At intermediate supercurrent, both order parameter components are finite at zero temperature and their transitions split (e.g. Fig.~\ref{fig4} (a)). In principle, the onset of the secondary order parameter component shall manifest as a change in the superfluid density or the magnetic penetration depth, which can be detected by scanning SQUID experiments~\cite{Mueller:23}. The second transition shall also naturally emerge as a specific heat anomaly, as we numerically verify in this section. We evaluate the specific heat according to the thermodynamic relation $C_{V}(T)=T\pd S(T)/\pd T$. Here, $S(T)$ represents the temperature-dependent entropy of the system, which can be derived from the partition function as follows:
\begin{align}\label{entropy1}
    S(T)=-\frac{\pd G}{\pd T},\quad G=-T\ln \tr (e^{-\beta \hH})
\end{align}
Given the Hamiltonian \eqref{mf_hamiltonian}, the definition \eqref{entropy1} yields the following expression for $S(\bq, T)$:
\begin{align}
    &S(\bq, T) \non \\
    = &-\sum_{\bk,\zeta=\pm}\big\{[1-f(E_{\bk,\zeta}(\bq))]\ln(1-f(E_{\bk,\zeta}(\bq)))\non \\
    &+f(E_{\bk,\zeta}(\bq))\ln f(E_{\bk,\zeta}(\bq))\big\}
\end{align}
where $f(E)=1/(e^{\beta E}+1)$ denotes the Fermi-Dirac distribution function and $E_{\bk,\pm}(\bq)$ has been defined in Eq. \eqref{energy}. Figure~\ref{fig4} (b) presents the temperature dependence of the specific heat for scenarios without and with supercurrent. In the latter scenario, a smaller but genuine anomaly emerges at the onset temperature of the secondary superconducting component shown in Fig.~\ref{fig4} (a).

In practice, specific heat measurement of a sample in a setup with externally applied current may be challenging. In particular, as the sample is thermally connected to external heat source via the electrodes through which the external current is applied, heat exchange near the electrode contacts cannot be circumvented. It is thus necessary to minimize the thermal contact by using, for example, thin and long gold or platinum wires for the electrodes. In addition, above the superconducting transition, the resistive sample is also heated on its own by the application of external current. To avoid this complication, it is better to perform the measurement from low to high temperatures. An alternative approach which can avoid the thermal contact issue is by using ring-shape samples. In this setup, supercurrent can be induced by threading a flux through the ring. In short, we expect that specific heat measurement in the presence of supercurrent should be feasible with deliberate experimental design. 

\subsection{Edge modes}
\label{subsec4}
The topologically non-trivial \chiralp state supports gapless chiral edge modes at the boundaries. This section focuses on how these edge modes are affected under the influence of the supercurrent. To analyze the behavior of the edge modes, we first take a continuum model and solve for the edge states at an open boundary parallel to $x$-axis. In this geometry, the $y$-component of the wavevectors are no longer good quantum numbers. Making the substitution $k_y \rightarrow -i\partial_y$, the Hamiltonian given in \eqref{BdG_matrix} becomes
\begin{align}\label{bdg_real}
    \begin{pmatrix}
    \frac{1}{2m}\big[(k_x+q_x)^2 - \pd_y^2 \big]-\mu & \frac{\Delta_x}{k_F}k_x + \frac{\Delta_y}{k_F}\pd_y\\
        \frac{\Delta_x}{k_F}k_x - \frac{\Delta_y}{k_F} \pd_y &   -\frac{1}{2m}\big[(k_x-q_x)^2 - \pd_y ^2\big]+\mu
    \end{pmatrix}
\end{align}
In the case of small supercurrent, i.e. $q_x/m \ll v_F$, we can simplify the Hamiltonian by neglecting the second-order term $q_x^2$. This allows us to rewrite the Hamiltonian as two distinct parts:
\begin{align}
    & \mathcal{H}_{k_x}(q_x) = \mathcal{H}_{1,k_x} + \frac{k_x}{m}q_x \sigma_0  \label{bdg_q1} ~~\text{with}\\
   & \mathcal{H}_{1,k_x} = \bigg[\frac{1}{2m}(k_x^2-\pd_y^2)-\mu\bigg]\sigma_3 + \frac{\Delta_x}{k_F}k_x\sigma_1 + \frac{i\Delta_y}{k_F}\pd_y \sigma_2
\end{align}
where $\sigma_{\mu}$ denotes the Pauli matrix. 

We note that $\mathcal{H}_{1,k_x}$ remains independent of $q_x$ and represents the original Hamiltonian without supercurrent, whose solutions of edge modes are well-documented in previous literature. The chiral edge dispersion is given by 
\begin{align}
    E_{\mathrm{edge},k_x}(q_x=0)=\frac{\Delta_x}{k_F}k_x
\end{align}
These states propagate unidirectionally at the boundary. For the time-reversed $p_x-ip_y$ state, the above dispersion acquires a minus sign and the propagation also switches direction. At finite $q_x$, we see from \eqref{bdg_q1} that the slope of the edge dispersion is changed by $q_x/m$,
\begin{align}
	E_{\mathrm{edge},k_x}(q_x) = \bigg( \frac{\Delta_x}{k_F} + \frac{q_x}{m}\bigg)k_x
\label{eq:edgeQ}
\end{align}
Reversing the supercurrent leads to an opposite change of slope, as depicted in the right panel of Fig.~\ref{fig0}. 

\begin{figure}
  \centering
  \includegraphics[width=1\linewidth]{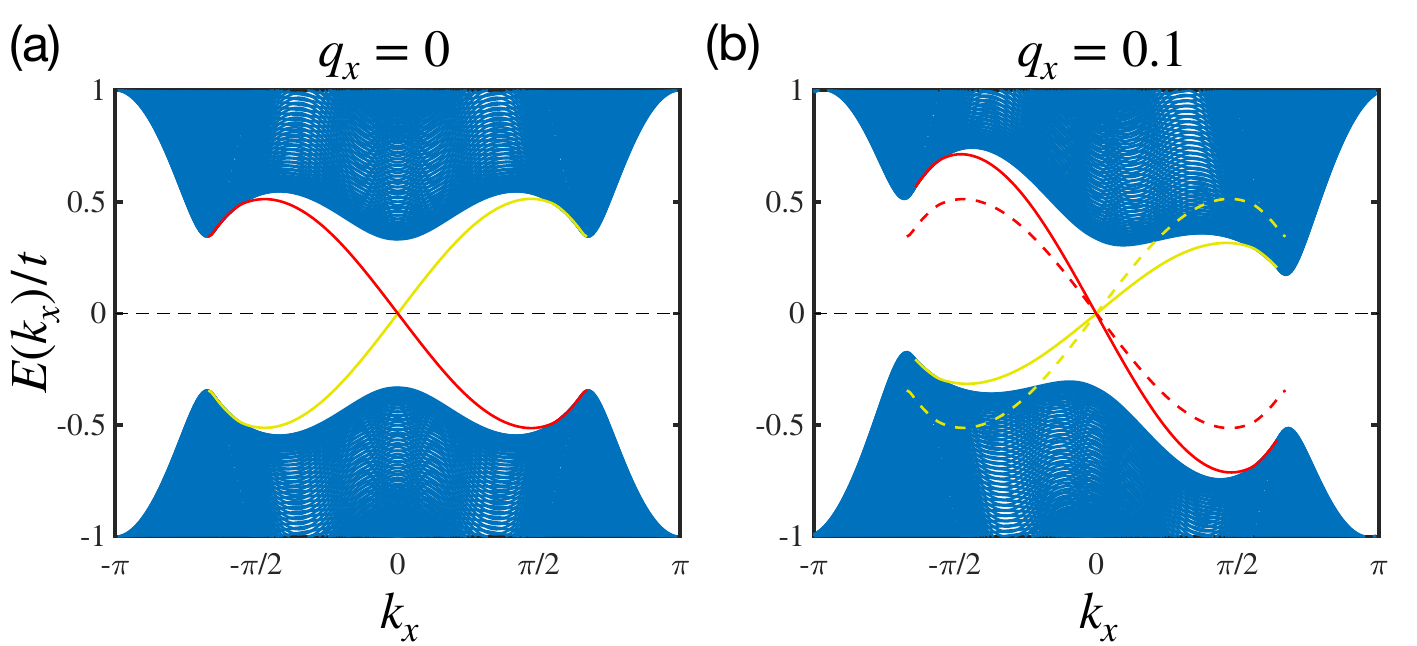}
  \caption{Dispersion of chiral p-wave models in a stripe geometry with periodic boundary conduction in $x$ direction, (a) without supercurrent, i.e. $q_x=0$ and (b) with finite supercurrent given by $q_x=0.1$. The red and green solid curves represent the chiral edge dispersions at opposite edges. In panel (b), the red and green dashed curves depict the chiral edge dispersion for the case when $q_x=0$. In these calculations, we take $\mu = 1$ and the interaction strength is $\widetilde{U}=-2.3 $.}
  \label{fig5}
\end{figure}

\begin{figure}[h]
  \centering
  \includegraphics[width=0.8\linewidth]{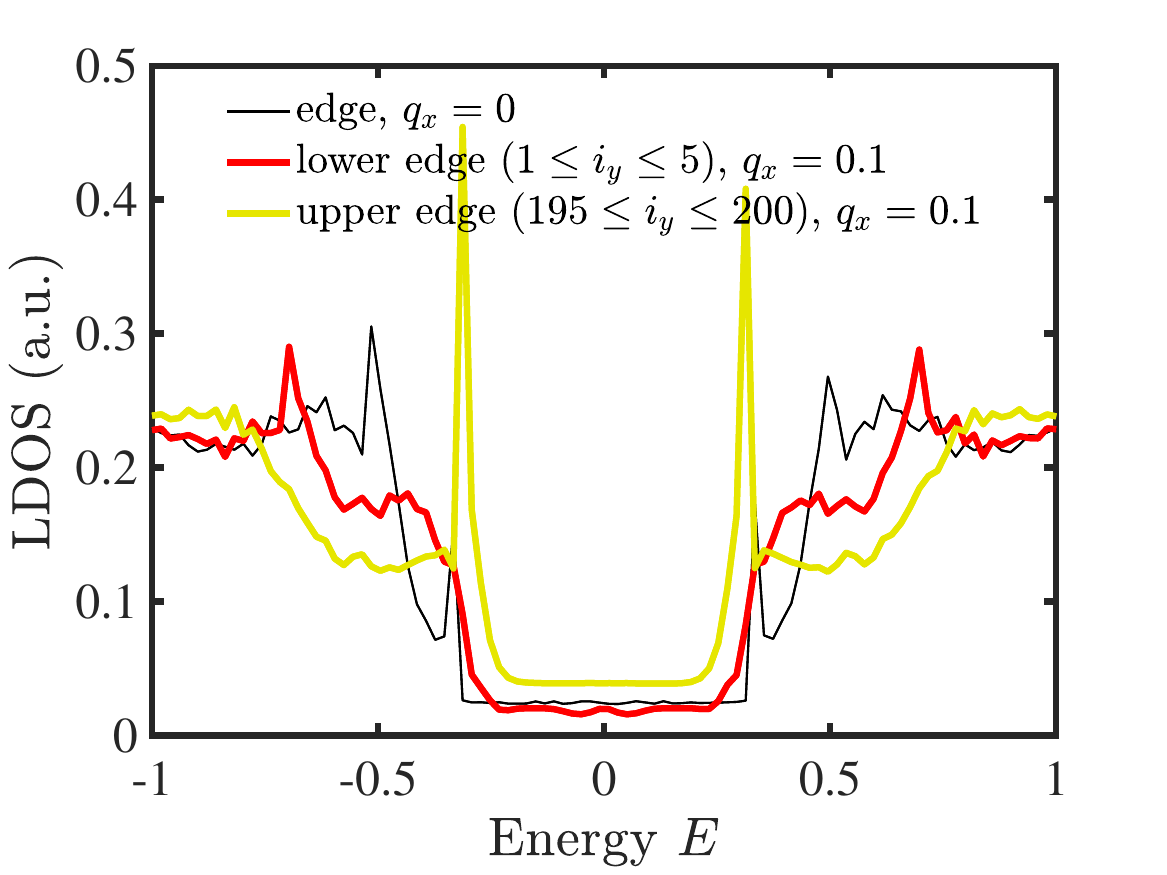}
  \caption{The averaged LDOS of five sites near two opposite edges in the presence of a supercurrent with $q_x=0.1$. The upper and lower edged are as illustrated in Fig. \ref{fig0}. At one edge, the applied supercurrent is parallel to the unidirectional propagation of chiral edge modes, and at another edge it is anti-parallel. Note that the lower edge result is equivalent to the upper edge one in the presence of an opposite supercurrent. For comparison, the $q_x=0$ curve shows the edge tunneling spectrum for $q_x=0$. The system consists of $N_y=200$ lattice sites in the $y$ direction, and the parameters used are the same as in Fig. \ref{fig5} (b).}
  \label{fig6}
\end{figure}

To verify the above semiclassical analysis, we perform numerical simulation of our square lattice model in a stripe geometry, with open boundaries in the $y$-direction and periodic boundary condition in the $x$-direction. For each supercurrent $q_x$, the Hamiltonian can be written as
\begin{align}
    \hH &= \sum_{k_x,i_y}[2t\cos (k_x+q_x)-\mu]\hcd_{k_x,i_y}\hc_{k_x+q_x,i_y}\non \\
    &+\sum_{k_x,i_y}(t\hcd_{k_x+q_x,i_y+1}\hc_{k_x+q_x,i_y}+\HC)\non \\
    &+\sum_{k_x,i_y}(\Delta_x \sin k_x\hcd_{k_x+q_x,i_y}\hcd_{-k_x+q_x,i_y}\non \\
    &+\Delta_y\hcd_{k_x+q_x,i_y+1}\hcd_{-k_x+q_x,i_y}\non \\
    &-\Delta_y\hcd_{k_x+q_x,i_y-1}\hcd_{-k_x+q_x,i_y}+\HC)
\end{align}
Figure \ref{fig5} compares the low energy spectrum of the above model for zero and finite $q_x$. As the chiral edge modes at the two open boundaries propagate in opposite direction, the supercurrent alters the slope of the two edge dispersions in opposite manner (Fig.~\ref{fig5} (b)), which is consistent with the above semiclassical expectation. Note that signatures of edge modes in \SRO~have been reported in some tunneling spectroscopic studies~\cite{Kashiwava2011, Ying2012}, although those data cannot confirm whether the pairing is chiral in nature. 

The peculiar response of chiral edge modes to opposite supercurrents naturally manifests as distinct changes in the edge tunneling spectrum. Figure \ref{fig6} shows the local density of states averaged over multiple sites in the vicinity of the two edges in the presence and in the absence of a supercurrent. Note that the lower edge result is equivalent to the upper edge one but with the supercurrent reversed. The difference is clearly discernible. In particular, compared to the $q_x=0$ case, the in-gap spectrum is shifted upward or downward, depending on the sign of $q_x$. This is consistent with the change of the slope of the edge dispersion, as described in \eqref{eq:edgeQ}. Some additional differences, which are model-dependent, can also be seen in the spectrum above the gap edge. The above results could provide a diagnosis for the chiral pairing, possibly more efficient than the experimental attempts to probe the spontaneous chiral edge current. This is because the spontaneous current in real samples could be rather small, hence the magnetic field generated by the spontaneous current could potentially be too weak to register a signal in actual experiments~\cite{Kirtley2007, Hicks2010, Curran2014}. 

In fact, the asymmetric supercurrent-induced corrections to the local density of states is not limited to sample edges. It is also expected in the vicinity of impurities, dislocations, or any other forms of crystalline defects around which spontaneous current may emerge (sketched on the left panel of Fig.~\ref{fig0}).

\section{Summary and further remarks}
\label{sec3}
In this study, we proposed to examine the possible chiral superconductivity in \SRO~and other materials through the application of a supercurrent. Akin to the effect of in-plane uniaxial strains and in-plane magnetic fields, supercurrent generically lifts the degeneracy of the two chiral superconducting order parameter components, giving rise to two superconducting transitions with observable thermodynamic and magnetic signatures. Moreover, opposite supercurrents incur distinct modifications to the unidirectionally propagating chiral edge modes and their tunneling spectra. We discussed possible measurements to detect these supercurrent-induced changes. While these results were obtained from single-band calculations, the conclusions hold for more general multiband systems. 

Note that our calculations have not accounted for the vortices generated by the supercurrent. Nonetheless, the conclusions are expected to hold even when those physics are considered. For example, while vortices drifting in supercurrent may give rise to fluctuating edge tunneling spectrum, the time-averaged spectra obtained under opposite supercurrents will likely still show discernible difference. We also take note of the stable tunneling spectra obtained in a conventional superconductor subject to a bias supercurrent~\cite{Anthore:03}. Our study shall also apply to the chiral $d_{xz}+id_{yz}$ state and the nonunitary mixed helical p-wave pairings~\cite{Huang:21}. The latter essentially consists of two copies of chiral p-wave pairing with opposite chirality and unequal gap amplitudes (thus referred to as chiral-like states). 

The subgap edge modes and spontaneous edge current may also appear in nonchiral but TRSB multi-component superconducting states. In fact, some of such states have been implicated in recent ultrasound measurements~\cite{ultrasound1,ultrasound2}. The same boundary phenomenology, including the supercurrent induced effects discussed above, may also occur at certain boundaries such states. For example, they can appear in $s+id_{xy}$ and $d_{x^2-y^2}+ig_{xy(x^2-y^2)}$ states on a square lattice at edges parallel to the $x$- or $y$-direction, but not at edges parallel to diagonals of the lattice~\cite{Furusaki:01}. Thus the dependence on edge orientation may help to distinguish chiral and nonchiral TRSB states, in addition to the polar Kerr effect measurement~\cite{Huang2021CPB,Huang:21}.

\section{Acknowledgements} 
We acknowledge many helpful comments and suggestions by Ying Liu, as well as multiple discussions with Wen Sun and Yongkang Luo. This work is supported by NSFC under Grants No.~11904155 and No.~12374042, the Guangdong Provincial Key Laboratory under Grant No.~2019B121203002, the Guangdong Science and Technology Department under Grant 2022A1515011948, a Shenzhen Science and Technology Program (Grant No. KQTD20200820113010023), and the Pingshan District Innovation Platform Project of Shenzhen Hi-tech Zone Development Special Plan in 2022 (Grant No. 29853M-KCJ-2023-002-01). Computing resources are provided by the Center for Computational Science and Engineering at Southern University of Science and Technology.


\begin{thebibliography}{0}
\bibitem{Luke1998} G. M. Luke, Y. Fudamoto, K. M. Kojima, M. I. Larkin, J. Merrin, B. Nachumi, Y. J. Uemura, Y. Maeno, Z. Q. Mao, Y. Mori, H. Nakamura, and M. Sigrist. Time-reversal symmetry-breaking superconductivity in Sr$_2$RuO$_4$.  \href{https://doi.org/10.1038/29038}{Nature {\bf 394}, 558–561 (1998)}.
%
\bibitem{Xia2006} J. Xia, Y. Maeno, P. T. Beyersdorf, M. M. Fejer, and A. Kapitulnik. High Resolution Polar Kerr Effect Measurements of 
Sr$_2$RuO$_4$: Evidence for Broken Time-Reversal Symmetry in the Superconducting State. \href{https://doi.org/10.1103/PhysRevLett.97.167002}{Phys. Rev. Lett. {\bf 97}, 167002 (2006)}.

\bibitem{Luke1993} G. M. Luke, A. Keren, L. P. Le, W. D. Wu, Y. J. Uemura, D. A. Bonn, L. Taillefer, and J. D. Garrett. Muon spin relaxation in UPt$_3$. \href{https://doi.org/10.1103/PhysRevLett.71.1466}{Phys. Rev. Lett. {\bf 71}, 1466 (1993)}.

\bibitem{Schemm2014} E. R. Schemm , W. J. Gannon, C. M. Wishne, W. P. Halperin, and A. Kapitulnik. Observation of broken time-reversal symmetry in the heavy-fermion superconductor UPt$_3$. \href{https://doi.org/10.1126/science.1248552}{Science {\bf 345}, 190-193(2014)}.

\bibitem{Mac1988} D. E. MacLaughlin, D. W. Cooke, R. H. Heffner, R. L. Hutson, M. W. McElfresh, M. E. Schillaci, H. D. Rempp, J. L. Smith, J. O. Willis, E. Zirngiebl, C. Boekema, R. L. Lichti, and J. Oostens. Muon spin rotation and magnetic order in the heavy-fermion compound 
URu$_2$Si$_2$. \href{https://doi.org/10.1103/PhysRevB.37.3153}{Phys. Rev. B {\bf 37}, 3153 (1988)}.

\bibitem{Schemm2015} E. R. Schemm, R. E. Baumbach, P. H. Tobash, F. Ronning, E. D. Bauer, and A. Kapitulnik. Evidence for broken time-reversal symmetry in the superconducting phase of URu$_2$Si$_2$. \href{https://doi.org/10.1103/PhysRevB.91.140506}{Phys. Rev. B {\bf 91}, 140506(R) (2015)}.

\bibitem{Maeno1994} Y. Maeno, H. Hashimoto, K. Yoshida, S. Nishizaki, T. Fujita, J.
G. Bednorz, F. Lichtenberg. Superconductivity in a layered perovskite without copper. \href{https://doi.org/10.1038/372532a0}{Nature (London) {\bf 372}, 532 (1994)}.

\bibitem{Maeno2001} Y. Maeno, T. M. Rice, and M. Sigrist. The Intriguing Superconductivity of Strontium Ruthenate. \href{https://doi.org/10.1063/1.1349611}{Phys. Today {\bf 54} (1), 42
(2001)}.

\bibitem{Maeno2003} A. P. Mackenzie and Y. Maeno. The superconductivity of 
Sr$_2$RuO$_4$ and the physics of spin-triplet pairing. \href{https://doi.org/10.1103/RevModPhys.75.657}{Rev. Mod. Phys. {\bf 75}, 657 (2003)}.

\bibitem{Kallin2009} C. Kallin and A. J. Berlinsky. Is Sr$_2$RuO$_4$ a chiral p-wave superconductor? \href{http://doi.org/10.1088/0953-8984/21/16/164210}{J. Phys.: Condens. Matter {\bf 21}, 164210 (2009)}. 

\bibitem{Kallin2012} C. Kallin. Chiral p-wave order in Sr$_2$RuO$_4$. \href{http://doi.org/10.1088/0034-4885/75/4/042501}{Rep. Prog. Phys. {\bf 75} 042501 (2012)}.

\bibitem{Maeno2012} Y. Maeno, S. Kittaka, T. Nomura, S. Yonezawa, and K. Ishida. Evaluation of Spin-Triplet Superconductivity in Sr$_2$RuO$_4$. \href{https://doi.org/10.1143/JPSJ.81.011009}{J. Phys. Soc. Jpn. {\bf 81}, 011009 (2012)}.

\bibitem{Liu2015} Y. Liu and Z. Q. Mao. Unconventional superconductivity in Sr$_2$RuO$_4$. \href{https://doi.org/10.1016/j.physc.2015.02.039}{Physica C: Superconductivity and its
Application, {\bf 514}, 339 (2015)}.

\bibitem{Kallin2016} C. Kallin and J. Berlinsky. Chiral superconductors. \href{https://doi.org/10.1088/0034-4885/79/5/054502}{Rep. Prog. Phys. {\bf 79},  054502 (2016)}.

\bibitem{Mac2017} A. P. Mackenzie, T. Scaffidi, C. W. Hicks and Y. Maeno. Even odder after twenty-three years: the superconducting order parameter puzzle of Sr$_2$RuO$_4$.  \href{https://doi.org/10.1038/s41535-017-0045-4}{npj Quant Mater {\bf 2}, 40 (2017)}.

\bibitem{Huang2021CPB} W. Huang. A review of some new perspectives on the theory of superconducting Sr$_2$RuO$_4$. \href{https://doi.org/10.1088/1674-1056/ac2488}{Chin. Phys. B, {\bf 30}, 107403 (2021)}.

\bibitem{Leggett2021} A. J. Leggett, Y. Liu. Symmetry Properties of Superconducting Order Parameter in Sr$_2$RuO$_4$.  \href{https://doi.org/10.1007/s10948-020-05717-6}{J Supercond Nov Magn {\bf 34}, 1647–1673 (2021)}. 

\bibitem{Ishida1998} K. Ishida, H. Mukuda, Y. Kitaoka, K. Asayama, Z. Q. Mao, Y. Mori and Y. Maeno. Spin-triplet superconductivity in Sr$_2$RuO$_4$ identified by 17O Knight shift. \href{https://doi.org/10.1038/25315}{Nature {\bf 396}, 658–660 (1998)}.


\bibitem{Duffy2000} J. A. Duffy, S. M. Hayden, Y. Maeno, Z. Mao, J. Kulda, and G. J. McIntyre. Polarized-Neutron Scattering Study of the Cooper-Pair Moment in Sr$_2$RuO$_4$. \href{https://doi.org/10.1103/PhysRevLett.85.5412}{Phys. Rev. Lett. {\bf 85}, 5412 (2000)}.

\bibitem{Liu2004} K. D. Nelson, Z. Q. Mao, Y. Maeno, Y. Liu. Odd-Parity Superconductivity in Sr$_2$RuO$_4$. \href{https://doi.org/10.1126/science.1103881}{Science {\bf 306}, 1151-1154(2004)}.


\bibitem{Pus2019} A. Pustogow, Yongkang Luo, A. Chronister, Y.-S. Su, D. A. Sokolov, F. Jerzembeck, A. P. Mackenzie, C. W. Hicks, N. Kikugawa, S. Raghu, E. D. Bauer and S. E. Brown. Constraints on the superconducting order parameter in Sr$_2$RuO$_4$ from oxygen-17 nuclear magnetic resonance. \href{https://doi.org/10.1038/s41586-019-1596-2}{Nature {\bf 574}, 72–75 (2019)}.

\bibitem{Chronister2021} A. Chronister, A. Pustogow, N. Kikugawa, D. A. Sokolov, F. Jerzembeck, C. W. Hicks, A. P. Mackenzie, E. D. Bauer, and S. E. Brown. Evidence for even parity unconventional superconductivity in Sr$_2$RuO$_4$.  \href{https://doi.org/10.1073/pnas.2025313118}{Proc. Natl. Acad. Sci. (USA) {\bf 118}, e2025313118 (2021)}.

\bibitem{exp1} E. Hassinger, P. Bourgeois-Hope, H. Taniguchi, S. Ren\'{e} de Cotret, G. Grissonnanche, M. S. Anwar, Y. Maeno, N. Doiron-Leyraud, and L. Taillefer. Vertical Line Nodes in the Superconducting Gap Structure of Sr$_2$RuO$_4$. \href{https://doi.org/10.1103/PhysRevX.7.011032}{Phys. Rev. X {\bf 7}, 011032 (2017)}.

\bibitem{exp2} V. Sunko, E. A. Morales, I. Marković, M. E. Barber, D. Milosavljević, F. Mazzola, D. A. Sokolov, N. Kikugawa, C. Cacho, P. Dudin, H. Rosner, C. W. Hicks, P. D. C. King and A. P. Mackenzie. Direct observation of a uniaxial stress-driven Lifshitz transition in Sr$_2$RuO$_4$. \href{https://doi.org/10.1038/s41535-019-0185-9}{npj Quantum Materials {\bf 4}, 46 (2019)}.

\bibitem{exp3} Y. S. Li, M. Garst, J. Schmalian, S. Ghosh, N. Kikugawa, D. A. Sokolov, C. W. Hicks, F. Jerzembeck, M. S. Ikeda, Z. Hu, B. J. Ramshaw, A. W. Rost, M. Nicklas and A. P. Mackenzie. Elastocaloric determination of the phase diagram of Sr$_2$RuO$_4$. \href{https://doi.org/10.1038/s41586-022-04820-z}{Nature {\bf 607}, 276 (2022)}.

\bibitem{ultrasound1}  S. Benhabib, C. Lupien, I. Paul, L. Berges, M. Dion, M. Nardone, A. Zitouni, Z. Q. Mao, Y. Maeno, A. Georges, L. Taillefer and C. Proust. Ultrasound evidence for a two-component superconducting order parameter in Sr$_2$RuO$_4$. \href{https://doi.org/10.1038/s41567-020-1033-3}{Nature Physics {\bf 17}, 194 (2021)}.

\bibitem{ultrasound2}  S. Ghosh, A. Shekhter, F. Jerzembeck, N. Kikugawa, D. A. Sokolov, M. Brando, A. P. Mackenzie, C. W. Hicks and B. J. Ramshaw. Thermodynamic evidence for a two-component superconducting order parameter in Sr$_2$RuO$_4$. \href{https://doi.org/10.1038/s41567-020-1032-4}{Nature Physics {\bf 17}, 199 (2021)}.

\bibitem{ZhangJL:20} J. L. Zhang, Y. Li, W. Huang, and F. C. Zhang. Hidden anomalous Hall effect in 
Sr$_2$RuO$_4$ with chiral superconductivity dominated by the Ru $d_{xy}$ orbital. \href{https://doi.org/10.1103/PhysRevB.102.180509}{Phys. Rev. B {\bf 102}, 180509(R) (2020)}.  

\bibitem{LiuHT:2023} H. T. Liu, W. Chen, and W. Huang. Impact of random impurities on the anomalous Hall effect in chiral superconductors.  \href{https://journals.aps.org/prb/abstract/10.1103/PhysRevB.107.224517}{Phys. Rev. B {\bf 107}, 224517 (2023)}.

\bibitem{Kivelson2020} S. A. Kivelson, A. C. Yuan, B. Ramshaw and R. Thomale. A proposal for reconciling diverse experiments on the superconducting state in Sr$_2$RuO$_4$. \href{https://doi.org/10.1038/s41535-020-0245-1}{npj Quantum Mater. {\bf 5}, 43 (2020)}. 

\bibitem{Sigrist1991} M. Sigrist and K. Ueda. Phenomenological theory of unconventional superconductivity. \href{https://doi.org/10.1103/RevModPhys.63.239}{Rev. Mod. Phys. {\bf 63}, 239–311 (1991)}.

\bibitem{Hicks2014} C. W. Hicks, D. O. Brodsky, E. A. Yelland, A. S. Gibbs, J. A. N. Bruin, M. E. Barber, S. D. Edkins, K. Nishimura, S. Yonezawa, Y. Maeno, and A. P. Mackenzie. Strong Increase of Tc of Sr$_2$RuO$_4$ Under Both Tensile and Compressive Strain.  \href{https://doi.org/10.1126/science.1248292}{Science {\bf 344}, 283-285 (2014)}.

\bibitem{Grinenko2021} V. Grinenko, S. Ghosh, R. Sarkar, J. C. Orain, A. Nikitin, M. Elender, D. Das, Z. Guguchia, F. Brückner, M. E. Barber, J. Park, N. Kikugawa, D. A. Sokolov, J. S. Bobowski, T. Miyoshi, Y. Maeno, A. P. Mackenzie, H. Luetkens, C. W. Hicks, and H.-H. Klauss. Split superconducting and time-reversal symmetry-breaking transitions in Sr$_2$RuO$_4$ under stress. \href{https://doi.org/10.1038/s41567-021-01182-7}{Nat. Phys. {\bf 17}, 748–754 (2021)}.

\bibitem{Li2021} Y. S. Li, N. Kikugawa, D. A. Sokolov, F. Jerzembeck, A. S. Gibbs, Y. Maeno, C. W. Hicks, J. Schmalian, M. Nicklas, and A. P. Mackenzie. High-sensitivity heat-capacity measurements on Sr$_2$RuO$_4$ under uniaxial pressure. \href{https://doi.org/10.1073/pnas.2020492118}{Proc. Natl. Acad. Sci. USA {\bf 118}, e2020492118 (2021)}.

\bibitem{Mueller:23} E. Mueller, Y. Iguchi, C. Watson, C. Hicks, Y. Maeno, K. Moler. Constraints on a split superconducting transition under uniaxial strain in Sr$_2$RuO$_4$ from scanning SQUID microscopy. \href{https://doi.org/10.48550/arXiv.2306.13737}{arXiv:2306.13737}. 



\bibitem{Steppke2017} A. Steppke, L. Zhao, M. E. Barber, T. Scaffidi, F. Jerzembeck, H. Bosner, A. S. Gibbs, Y. Maeno, S. H. Simon, A. P. Mackenzie and C. W. Hicks. Strong peak in Tc of Sr$_2$RuO$_4$ under uniaxial pressure. \href{https://doi.org/10.1126/science.aaf9398}{Science {\bf 355}, eaaf9398 (2017)}. 

\bibitem{Watson2018} C. A. Watson, A. S. Gibbs, A. P. Mackenzie, C. W. Hicks, and K. A. Moler. Micron-scale measurements of low anisotropic strain response of local $T_c$ in Sr$_2$RuO$_4$. \href{https://doi.org/10.1103/PhysRevB.98.094521}{Phys. Rev. B {\bf 98}, 094521 (2018)}.

\bibitem{Agterberg:98} D.F. Agterberg. Vortex Lattice Structures of Sr$_2$RuO$_4$. \href{https://doi.org/10.1103/PhysRevLett.80.5184}{Phys. Rev. Lett. {\bf 80}, 5184 (1998)}. 

\bibitem{Kaur:05} R.P. Kaur, D.F. Agterberg, H. Kusunose. Quasiclassical determination of the in-plane magnetic field phase diagram of superconducting Sr$_2$RuO$_4$. \href{https://doi.org/10.1103/PhysRevB.72.144528}{Phys. Rev. B {\bf 72}, 144528 (2005)}. 

\bibitem{Yonezawa2014} S. Yonezawa, T. Kajikawa, and Y. Maeno. Specific-Heat Evidence of the First-Order Superconducting Transition in Sr$_2$RuO$_4$. \href{https://doi.org/10.7566/JPSJ.83.083706}{	J. Phys. Soc. Jpn. {\bf 83}, 083706 (2014)}.

\bibitem{Anthore:03} A. Anthore, H. Pothier, and D. Esteve. Density of States in a Superconductor Carrying a Supercurrent.  \href{https://journals.aps.org/prl/abstract/10.1103/PhysRevLett.90.127001}{Phys. Rev. Lett. {\bf 90}, 127001 (2003)}. 

\bibitem{Zhang2004} D. Zhang, C. S. Ting, and C. R. Hu. Conductance characteristics between a normal metal and a clean superconductor carrying a supercurrent. \href{https://doi.org/10.1103/PhysRevB.70.172508}{Phys. Rev. B 70, 172508 (2004)}.

\bibitem{Yer2022} Y. Yerin, S. L. Drechsler, M. Cuoco, and C. Petrillo. Magneto-topological transitions in multicomponent superconductors. \href{https://doi.org/10.1103/PhysRevB.106.054517}{Phys. Rev. B {\bf 106}, 054517 (2022)}.

\bibitem{Yer2023} Y. Yerin, S. L. Drechsler, M. Cuoco and C. Petrillo. Multiple-q current states in a multicomponent superconducting channel. \href{http://iopscience.iop.org/article/10.1088/1361-648X/acf42d}{J. Phys. Condens., Accepted Manuscript (2023)}

\bibitem{FF1964} P. Fulde and R. A. Ferrell. Superconductivity in a Strong Spin-Exchange Field.  \href{https://doi.org/10.1103/PhysRev.135.A550}{Phys. Rev. {\bf 135}, A550 (1964)}.

\bibitem{Stone2004} M. Stone and R. Roy. Edge modes, edge currents, and gauge invariance in $p_x+ip_y$ superfluids and superconductors. \href{https://doi.org/10.1103/PhysRevB.69.184511}{Phys. Rev. B {\bf 69}, 184511 (2004)}.

\bibitem{Huang2015} W. Huang, S. Lederer, E. Taylor, and C. Kallin. Nontopological nature of the edge current in a chiral $p$-wave superconductor.  \href{https://doi.org/10.1103/PhysRevB.91.094507}{Phys. Rev. B {\bf 91}, 094507an (2015)}.

\bibitem{Tada2015} Y. Tada. Equilibrium surface current and role of U(1) symmetry: Sum rule and surface perturbations.  \href{https://doi.org/10.1103/PhysRevB.92.104502}{Phys. Rev. B 92, 104502 (2015)}.

\bibitem{Imai201213} Y. Imai, K. Wakabayashi, and M. Sigrist. Properties of edge states in a spin-triplet two-band superconductor.  \href{https://doi.org/10.1103/PhysRevB.85.174532}{Phys. Rev. B {\bf 85},
174532 (2012)}; \href{https://doi.org/10.1103/PhysRevB.88.144503}{{\bf 88}, 144503 (2013)}.

\bibitem{Huang2014} W. Huang, E. Taylor, and C. Kallin. Vanishing edge currents in non-$p$-wave topological chiral superconductors.  \href{https://doi.org/10.1103/PhysRevB.90.224519}{Phys. Rev. B {\bf 90}, 224519 (2014)}.

\bibitem{Lederer2014} S. Lederer, W. Huang, E. Taylor, S. Raghu, and C. Kallin. Suppression of spontaneous currents in Sr$_2$RuO$_4$ by surface disorder. \href{https://doi.org/10.1103/PhysRevB.90.134521}{Phys.
Rev. B {\bf 90}, 134521 (2014)}.

\bibitem{Scaffidi2015} T. Scaffidi and S. H. Simon. Large Chern Number and Edge Currents in Sr$_2$RuO$_4$. \href{https://doi.org/10.1103/PhysRevLett.115.200402}{Phys. Rev. Lett. {\bf 115}, 087003 (2015)}.

\bibitem{Tada2015b} Y. Tada, W. Nie, and M. Oshikawa. Orbital Angular Momentum and Spectral Flow in Two-Dimensional Chiral Superfluids.  \href{https://doi.org/10.1103/PhysRevLett.114.195301}{Phys. Rev. Lett. {\bf 114}, 195301 (2015)}.

\bibitem{Kirtley2007} J. R. Kirtley, C. Kallin, C. W. Hicks, E. A. Kim, Y. Liu, K. A. Moler, Y. Maeno, and K. D. Nelson. Upper limit on spontaneous supercurrents in Sr$_2$RuO$_4$. \href{https://doi.org/10.1103/PhysRevB.76.014526}{Phys. Rev. B {\bf 76}, 014526 (2007)}.

\bibitem{Hicks2010} C. W. Hicks, J. R. Kirtley, T. M. Lippman, N. C. Koshnick, M. E. Huber, Y. Maeno, W. M. Yuhasz, M. B. Maple, and K. A. Moler. Limits on superconductivity-related magnetization in Sr$_2$RuO$_4$ and PrOs$_4$Sb$_{12}$ from scanning SQUID microscopy. \href{https://doi.org/10.1103/PhysRevB.81.214501}{Phys. Rev. B 81, 214501 (2010)}.

\bibitem{Curran2014} P. J. Curran, S. J. Bending, W. M. Desoky, A. S. Gibbs, S. L. Lee, and A. P. Mackenzie. Search for spontaneous edge currents and vortex imaging in Sr$_2$RuO$_4$ ~mesostructures. \href{https://doi.org/10.1103/PhysRevB.89.144504}{Phys. Rev. B {\bf 89}, 144504 (2014)}.


\bibitem{Tinkham2004} M. Tinkham, Introduction to Superconductivity. 2nd Edition, Dover Publication, Mineola, New York, 2004.

\bibitem{Volovik1993} G. E. Volovik. Superconductivity with lines of GAP nodes: density of states in the vortex. \href{https://ui.adsabs.harvard.edu/abs/1993JETPL..58..469V}{JETP Lett. {\bf 58}, 469 (1993)}.


\bibitem{Kashiwava2011} S. Kashiwaya, H. Kashiwaya, H. Kambara, T. Furuta, H. Yaguchi, Y. Tanaka, and Y. Maeno.Edge States of Sr$_2$RuO$_4$ Detected by In-Plane Tunneling Spectroscopy. \href{https://doi.org/10.1103/PhysRevLett.107.077003}{Phys. Rev. Lett. {\bf 107}, 077003 (2011)}.

\bibitem{Ying2012} Yiqun Alex Ying. Probing the enhanced superconductivity and chiral edge current in spin-triplet superconductor \SRO. \href{https://etda.libraries.psu.edu/files/final_submissions/7991}{Ph.D. dissertation in physics} at the Pennsylvania State University, 2012.

\bibitem{Huang:21} W. Huang and Z. Wang. Possibility of mixed helical p-wave pairings in \SRO. \href{https://doi.org/10.1103/PhysRevResearch.3.L042002}{Phys. Rev. Res. {\bf 3}, L042002 (2021)}. 

\bibitem{Furusaki:01} A. Furusaki, M. Matsumoto, and M. Sigrist. Spontaneous Hall effect in a chiral $p$-wave superconductor. \href{https://doi.org/10.1103/PhysRevB.64.054514}{Phys. Rev. B {\bf 64}, 054514 (2001)}. 

 
\end{thebibliography}
\end{document}